\documentclass[manuscript]{acmart}
\AtBeginDocument{%
  \providecommand\BibTeX{{%
    \normalfont B\kern-0.5em{\scshape i\kern-0.25em b}\kern-0.8em\TeX}}}

\setcopyright{acmcopyright}
\copyrightyear{2018}
\acmYear{2018}
\acmDOI{XXXXXXX.XXXXXXX}

\acmConference[Conference acronym 'XX]{Make sure to enter the correct
  conference title from your rights confirmation emai}{June 03--05,
  2018}{Woodstock, NY}
\acmPrice{15.00}
\acmISBN{978-1-4503-XXXX-X/18/06}

\usepackage{booktabs}
\usepackage{colortbl}

\usepackage[normalem]{ulem}
\useunder{\uline}{\ul}{}

\newcommand{\ejain}[1]{{\color{red}ejain:#1}\normalfont}

\usepackage{gensymb}

\begin{document}

\title{Transit drivers' reflections on the benefits and harms of eye tracking technology}

\author{Shaina Murphy}
\affiliation{%
  \institution{University of Florida}
  \city{Gainesville, Florida}
  \country{USA}}
\email{shainanmurphy@ufl.edu}

\author{Bryce Grame}
\affiliation{%
  \institution{University of Florida}
  \city{Gainesville, Florida}
  \country{USA}}
\email{brycegrame@ufl.edu}

\author{Ethan Smith}
\affiliation{%
  \institution{University of Florida}
  \city{Gainesville, Florida}
  \country{USA}}
\email{ethansmith@ufl.edu}

\author{Siva Srinivasan}
\affiliation{%
  \institution{University of Florida}
  \city{Gainesville, Florida}
  \country{USA}}
\email{siva@ce.ufl.edu}

\author{Eakta Jain}
\affiliation{%
  \institution{University of Florida}
  \city{Gainesville, Florida}
  \country{USA}}
\email{ejain@cise.ufl.edu}

\renewcommand{\shortauthors}{Murphy et al.}

\begin{abstract}
Eye tracking technology offers great potential for improving road safety. It is already being built into vehicles, namely cars and trucks. When this technology is integrated into transit service vehicles, employees, i.e., bus drivers, will be subject to being eye tracked on
their job. Although there is much research effort advancing algorithms for eye tracking in transportation, less is known about how end users perceive this technology, especially when interacting with it in an employer-mandated context. In this first study of its kind, we investigated transit bus operators’ perceptions of eye tracking technology. From a methodological perspective, we introduce a mixed methods approach where participants experience the technology first-hand and then reflect on their experience while viewing a playback of the recorded data. Thematic analysis of the interview transcripts reveals interesting potential uses of eye tracking in this work context and surfaces transit operators’ fears and concerns about this technology.
\end{abstract}




\received{20 February 2007}
\received[revised]{12 March 2009}
\received[accepted]{5 June 2009}

\maketitle
\section{Introduction}


Much research has looked into how eye tracking technology can be used for driver attention tracking \cite{bortkiewicz2019analysis},\cite{ahlstrom2018Stress},\cite{madlenak2023EyeTracking},\cite{raddaoui2020Evaluatinga}, advanced driver assistance \cite{gruchmann2021Bus},\cite{lotz2019Response},\cite{diederichs2020Adaptive}, fatigue and drowsiness tracking \cite{castritius2021Driver},\cite{mandal2017Detection},\cite{shekarisoleimanloo2022Association}, and distinguishing between experts and novices drivers \cite{ribeiro2021VR}. While there is tremendous promise in how eye tracking technology can contribute towards the aforementioned goals, there is a gap in the literature on how end user drivers perceive this technology. When end user perceptions have not been sought out, an otherwise well-intentioned technological improvement can, in practice, result in unanticipated or unintended consequences as well as lower acceptance and adoption of the technology within the specific work context. This was famously shown in the context of electronic logging devices mandated for long haul trucking \cite{levy2022Dataa}. 

There is a technological tension that exists between the possible societal improvements that can come about by employing new technologies and people’s initial resistance to these technologies, which motivates prototyping. As Astfalk and colleagues note, “Using physical prototypes that are not fully functional at an early stage of technology maturity could lead to lower resource consumption as customer feedback could be incorporated faster and acceptance could be established. However, previous research has indicated differences in consumer attitudes toward technologies before and after users have experienced prototypes first-hand”~\cite{ASTFALK2021100444}. This tension dissolves at the point of acceptance, when people begin to understand that a new technology's benefits (perceived, potential, or otherwise) outweigh their initial fears. In other words, soliciting end user perspectives can on the one hand surface potential problems that can be addressed at the design stage, and on the other hand, also reduce ``fear of the unknown'' which could improve acceptance of the technology when it is deployed. 

In this paper, we take a user-centered approach to understanding how eye tracking technology can be used in the transportation industry. We focus on the public transportation sector by examining how bus drivers perceive eye tracking. Prior to the study, we hypothesized that transit bus operators would perceive eye tracking as a means of monitoring. We also conjectured the transit bus operators could perceive eye tracking to be helpful in training, and that eye tracking glasses could be burdensome to the task of driving. To explore these hypotheses, we collaborated with a city bus transit agency to conduct a naturalistic field study with currently employed city bus operators. As part of the study, we provided them first-hand, real-world experience with eye tracking technology. We  reviewed the playback recorded by the eye tracker with each participant and queried their perceptions of the eye tracking technology within a semi-structured interview format derived from the technology acceptance model. Thematic analysis of participants’ answers generated insight into our initial conjectures from the perspective of eventual end users. 

 Our \textbf{contributions} to eye tracking literature are twofold: First, we have documented transit bus operators' perceptions of eye tracking technology in a naturalistic field study after both a first-hand, real-world experience with the technology and a viewing of the playback the eye tracker captured of the driver’s own driving. Secondly, we provide a methodology for mixed methods research that may be replicated for eliciting end user perceptions of eye tracking technology in future naturalistic field studies in different contexts.

\section{Background}

Though there has been much research on eye tracking in the transportation context~\cite{kapitaniak2015application}, there have been only a few studies that have focused on eye-tracking in a transit setting. Most studies relied on lab simulation, for example, using the existing surveillance camera in a bus as an eye-tracker~\cite{mandal2017Detection}, using dash-mounted eye-trackers fitted in replica transit bus cockpits to identify differences in visual strategies used by drivers in high-risk scenarios ~\cite{bortkiewicz2019analysis}, to gather data on stress, fatigue and glance behavior for developing driver state detection algorithms~\cite{ahlstrom2018Stress}, or using head-mounted eye tracking glasses to assess differences in visual strategy (e.g., number and duration of fixations) when bus drivers underwent two simulated rides on a city bus simulator, one into the city center and the other in the suburbs, each involving a simulated dangerous incident~\cite{kapitaniak2020visual}. All these studies used eye tracking technology as a measurement tool but none asked questions related to the participants’ (i.e., drivers) perceptions of the technology itself. The recorded eye tracking data were not replayed to the participants because participants' perceptions were not the focus of the research. 

Even in studies focused on other public transportation modes, perceptions about the eye tracking technology have not been a focus of the research; eye-tracking has primarily been used as a means of evaluation even when the studies have relied on qualitative research methods such as focus groups and case studies. For example, in the urban rail context, Rjabovs and colleagues conducted a case study with four drivers using eye tracking technology to investigate the impact of changing design elements such as mirrors, platforms, stopping position markers, and signal positions~\cite{rjabovs2019investigation}. Warchol-Jakubowska and colleagues collected eye tracking data while watching a tram driving simulation to compare novice and expert tram drivers~\cite{krejtz:etra:2023},
Kircher and colleagues~\cite{kircher2020effects} conducted eye tracking of truck drivers in a naturalistic field study: after an initial drive in an urban setting, inexperienced truck drivers underwent anticipatory driving training, followed by a subsequent drive along the same route. Using eye tracking, the study evaluated the impact of driver training and found increased monitoring of cyclists. 
Even when eye tracking studies have employed interviews, they have typically asked participants questions related to task performance to provide context for eye tracking data, but not to address technology perception issues. Literature that has studied technology perception indicates that truck drivers can have a positive attitude toward receiving feedback through technology, though not specifically eye tracking, for improving driving safety based on focus groups and surveys~\cite{huang2005feedback}. We note that the nature of work for long haul truckers is quite different compared to transit bus drivers. Bus drivers operate on city streets, typically involving congested intersections and traffic signals, while long haul truckers spend most of their time on highways at relatively higher speeds. Bus drivers also interact with the public and perform auxiliary activities such as helping people secure their wheelchairs and bikes or giving directions to tourists. 

Taken together, there are two knowledge gaps in the current literature: from a subject matter perspective, the gap is perception of eye tracking technology in the work context of transit driving; from a methodological perspective, the gap is how to solicit these perceptions for a technology that is commodity enough to be on employers' radars but not commodity enough that workers would have experienced it in their daily lives and be able to provide informed responses in surveys. 

\section{Methods}
Qualitative methods in the form of field observations and interviews have long been used to gauge and assess individual’s attitudes, beliefs, and experiences~\cite{esterberg2002Qualitative,saldana2023Coding}. Quantitative methods like the NASA Task Loading Index (TLX) have been used in an extensive body of literature to subjectively assess workload~\cite{TLX,hart2006NasaTask}. In this naturalistic field study, we adopted semi-structured interviews and subsequently qualitative coding of those interviews as the main source of evidence to investigate and understand transit drivers’ perceptions of eye tracking technology within their real-world work environment. A naturalistic field study was chosen as it would orient participants to real world use cases and additionally provide hands-on experience to participants who had never interacted with eye tracking prior to this study. 

\subsection{Field Experiment}
Close collaboration with a local transit agency was key to the field work. The local transit agency provided the study team with access to resources which included a liaison to assist with scheduling the study, a conference room to conduct interviews with participants, out-of-service buses to navigate the study route, and lastly man hours which allowed bus drivers who volunteered in their downtime to participate in the study. The study protocol was developed in collaboration with the transit agency and approved by the University of XX Institutional Review Board. 
Driving weather conditions across all trials ranged from mild rain to bright sun and no more than two trials were completed a day (morning and afternoon). 

\subsubsection{Protocol}
Every participant was explained the purpose of the study, walked through a printed map of the study bus route, and questions were addressed. The study consisted of three parts: an intake interview, hands-on experience with the eye tracking technology along the study bus route, and a post interview. 
In the first part, participants were asked how long they have driven for the transit agency, what they enjoy about driving buses, and what pain points they experience doing so. Participants and the research team then boarded the bus to begin the second part of the study. The participant performed a safety check of the vehicle. Then, a member of the research team assisted the participant with both wearing the eye tracking glasses and calibrating it. SensoMotoric Instruments (SMI) Eye Tracking Glasses 2 were used and after a 3-point calibration process the expected gaze position accuracy is 0.5\degree over all distances. These head-mounted eye tracking glasses report up to 120Hz gaze data and the camera feed from its scene camera.


Following a confirmation to depart the bus depot by the transit agency dispatch, the participant pulled out of the depot and began following the study route. After reaching the halfway point of the study route located at a nearby bus terminal, the participant parked the bus and the research team re-calibrated the eye tracking glasses if needed. The participant was then asked to proceed with the last half of the route and return to the bus depot. Participants were familiar with the area and only requested that the research team cue them with landmarks in order to traverse the study route. 

The third part of the study was conducted in the conference room. The participant reviewed the playback of the gaze overlay video exported by the eye tracker. 
The participant was asked their general thoughts on eye tracking technology, how easy it was to use, in what situations might eye tracking be useful, who they thought would benefit the most from eye tracking, and where might eye tracking be useful. The research team also asked their opinions on who might be most harmed by eye tracking and in what situations could eye tracking be harmful. Then the participant filled out the NASA TLX. Finally, the participant was asked if they learned anything new and was thanked for their participation. 



\subsubsection{Participants}
Ten bus drivers, five female and five male, volunteered to participate in the study. The small number was due to participant availability being limited to times when transit bus usage was low. Participant experience with the transit agency ranged from two to sixteen years, with a mean of 9.4 years (standard deviation = 5.4). 
Eligibility criteria for each participant included the ability to drive without correctional glasses to prevent calibration issues with the eye tracking glasses. All ten participants completed the study. 



\subsection{Qualitative Analysis}
The research team conducted a thematic analysis of interview artifacts according to the process described by Braun and Clarke \cite{braun2006Using}. Interview data was first transcribed using Otter.ai, then the research team refined the transcripts for accuracy. To begin coding, the research team familiarized itself with the interview data by reviewing all of the collected interview transcripts, recordings, and notes while penciling down recurring and interesting codes related to participant perceptions of eye tracking technology. Participants' perceived benefits and harms as well as perceived ease of use of eye tracking were the focus of this pass. On a second pass, an initial set of themes was developed by gathering similar, previously noted codes together. 

One researcher provided a blank table formatted with columns housing the list of all initial themes with their meanings and rows with participant excerpts to two other coders. All three coders then independently coded the set of $45$ excerpts into the initial theme they believed best fit each excerpt. Although Braun and Clarke state intercoder agreement measures are problematic \cite{oconnor2020Intercoder}, Fleiss's kappa statistic was calculated in SPSS for these initial themes to assess agreement between the three coders, $\kappa$ = 0.803. According to Landis and Koch's scale, this is substantial intercoder agreement \cite{landis1977Measurement}. 

Afterwards, the three coders engaged in group discussion amongst the research team which led to a final pass that further refined themes by getting rid of any redundancy, merging similar themes, and reviewing themes in relation to the dataset and our research questions. The resulting final thematic map of bus driver perception depicted in Table \ref{tab:tm}.

\subsection{Quantitative Analysis}
The Official NASA-TLX is a tool allowing users to perform subjective workload assessments on human-machine interface system operator. The tool was initially a paper and pencil questionnaire, developed by NASA Ames Research Center’s (ARC) Sandra Hart in the 1980s. Since then, it is broadly accepted as the  standard for many industries in measuring subjective workload. The TLX a multi-dimensional rating procedure to derive an overall workload score, which is based on a weighted average of ratings on six subscales: Mental Demand, Physical Demand, Temporal Demand, Performance, Effort, and Frustration~\cite{article}.
The TLX can be weighted, but for this study the raw TLX scores were taken without weighting~\cite{hancock1988human}.
The exact language of these questions is provided in the Appendix. 

\section{Results}
Five overarching themes were identified from our thematic analysis. 

\begin{table}[]
\caption{Thematic map of bus driver perceptions towards eye trackers within their work context.}
\label{tab:tm}
\begin{tabular}{|l|l|}
\hline
\multicolumn{1}{|c|}{Final Themes} &
  \multicolumn{1}{c|}{Examples} \\ \hline
\rowcolor[HTML]{C0C0C0} 
{\ul \textbf{Training}} &
   \\ \hline
\textbf{Self Assessment} &
  \textit{\begin{tabular}[c]{@{}l@{}}Participant 9: "…now I just wish everybody could wear these things\\ and just take a look at it once they get home and get situated, like.,\\ “Oh my god.” Because you’ve got most people like this {[}head \\ gesture with stiff neck{]} no eye movement, none of that." \\ {[}code: reviewing playback{]}\end{tabular}} \\ \hline
\textbf{Improved Training} &
  \textit{\begin{tabular}[c]{@{}l@{}}Participant 8: "And you use that in a training class of people who \\ are trying to get their license, it can be used as an example for new \\ drivers of:'this is what you need to be doing or not doing.'" \\ {[}code: training{]}\end{tabular}} \\ \hline
\rowcolor[HTML]{C0C0C0} 
{\ul \textbf{Tool for Communication}} &
   \\ \hline
\textbf{Empathy Towards Driver} &
  \textit{\begin{tabular}[c]{@{}l@{}}Participant 10: "The actual, any transit facility, for even for the boss, \\ they can see what a driver go through..." {[}code: empathy for driver{]}\end{tabular}} \\ \hline
\textbf{Roadway Improvement} &
  \textit{\begin{tabular}[c]{@{}l@{}}Participant 10: "Also showing like roadways like what could be \\ improved on the road. Like, wow, a big vehicle going through they \\ might need to expand the road or move a bicycle lane somewhere..." \\ {[}code: improve roads for bus{]}\end{tabular}} \\ \hline
\rowcolor[HTML]{C0C0C0} 
{\ul \textbf{Ease of Use}} &
   \\ \hline
\textbf{Outside Their Norm} &
  \textit{\begin{tabular}[c]{@{}l@{}}Participant 1: "I don’t think a transit driver cause some of us don't \\ wear glasses or shades, so it’d be kind of... (shrug)" \\ {[}code: no glasses{]}\end{tabular}} \\ \hline
\textbf{Comfortability} &
  \textit{\begin{tabular}[c]{@{}l@{}}Participant 9: "The glasses could be a little looser because I got a \\ fat face." {[}code: adjustments{]}\end{tabular}} \\ \hline
\textbf{Distraction} &
  \textit{\begin{tabular}[c]{@{}l@{}}Participant 10: "I wasn't irritated or stress. Only thing. Okay. I \\ would just give it a one because of the cord. Okay. And that's only \\ because when I was turning, I could feel it. It wasn’t stopping me \\ from turning but it was weird." {[}code: cord{]}\end{tabular}} \\ \hline
\rowcolor[HTML]{C0C0C0} 
{\ul \textbf{Accident Analysis}} &
   \\ \hline
\textbf{Accident Prevention} &
  \textit{\begin{tabular}[c]{@{}l@{}}Participant 10: "It can help in preventing accidents." \\ {[}code: prevents accidents{]}\end{tabular}} \\ \hline
\textbf{Accident Analysis} &
  \textit{\begin{tabular}[c]{@{}l@{}}Participant 2: "Actually seeing what a driver sees with safety you \\ know, if an accident occurred what was the driver actually looking \\ at before he made a certain turn if they did get an accident or \\ something like that." {[}code: fault in accident{]}\end{tabular}} \\ \hline
\rowcolor[HTML]{C0C0C0} 
{\color[HTML]{000000} {\ul \textbf{Fear of Replacement/Loss of Livelihood}}} &
  {\color[HTML]{000000} } \\ \hline
\textbf{Enabling Artificial Intelligence} &
  \textit{\begin{tabular}[c]{@{}l@{}}Participant 8: "I think the technology is good as long as you're not\\ going to use it to design an AI that replaces my job." \\ {[}code: enables AI{]}\end{tabular}} \\ \hline
\textbf{Use Against Driver} &
  \textit{\begin{tabular}[c]{@{}l@{}}Participant 2: "...it's recording everything you're looking at so it \\ could be used against you ... Court, got into an accident or \\ something." {[}code: use against driver{]}\end{tabular}} \\ \hline
\end{tabular}
\end{table}

\subsection{Theme 1: Training}
This theme deals with accident prevention through improved training for novice drivers and continuous training for current drivers. Participants generally brought up the potential for eye tracking technology as a novel tool for reflection and self-improvement. By reviewing the gaze replay of expert transit drivers, strategies for safe driving could be illustrated to novice drivers as part of formal training. They might even be able to view their own data and an expert's data on the same tricky intersection to compare and contrast. \textit{Participant 8: And you use that in a training class of people who are trying to get their license, it can be used as an example for new drivers of: “this is what you need to be doing or not doing.} Though training videos are used routinely, participants pointed out that they show what was happening on the road but not what the driver was paying attention to. Using gaze replays, trainees can see what the driver was attending to in addition to the action they took. \textit{Participant 10: Because you cannot just like, we have videos, when in training, we have videos of someone driving, but you can't tell what they look at. Yeah, you will say, so how can you say, where do like this driver on a video eyes go? Well, you still don't know what you you know what, like key points in the road or in a mirror you need to look at.} Another possible training use case would be for continuous improvement on the job. For example, a driver may review their own data from time to time. Alternately, a gaze replay might be included as part of a peer review process. All in all, eye tracking technology opens up the avenue of reflection-based training and improvement for accident prevention. 
\textit{Participant 7: To see what other drivers seeing, yeah, just to see what others they have not seen. We can learn from what they was looking at in a situation that we would be in also.}

\textbf{Implications} These findings suggest that it would be both easy and beneficial for driver training to incorporate gaze replays during early stage training and continuous on-the-job training. Research would be needed to quantify the improvement in driver performance as a result of this type of intervention in the training process. 

\subsection{Theme 2: Tool for Communication}
An interesting theme that emerged from participant interviews is how eye tracking technology can serve as a tool for commmunication between stakeholders at different levels of power and decision making positions. Reviewing the gaze replay, or using portions of it as a visual aid during discussions, allows for the creation of an ``empathy window'' through which managers and decision makers can experience the daily work of driving through the eyes of the transit driver.
\textit{Participant 10: The actual, any transit facility, for even for the boss, they can see what a driver go through. 
If you don't drive every day, you cannot make a route or you know, create a route, or tell the driver how to drive, or if you don't know or see what they doing or going through...Also showing like roadways like what could be improved on the road. Like, wow, a big vehicle going through they might, they might need to expand the road or, you know, move a bicycle lane somewhere, or like we was just talking about the crosswalks, you know, this is what a bus do every day. So maybe y’all might need to move this, or like just say if we had something to tell us that somebody you know walking across and it can cover our blind spots.}

\textbf{Implications} Eye tracking technology could be used to communicate why a certain intersection is challenging or why driving within a college campus poses challenges that are not typically considered while training novice drivers. A broader implication of this finding is that eye tracking data could similarly be used as a tool for communication in other contexts too, for example, interactions between buses and pedestrians in urban areas, the challenges of truck driving, the dangers of bicycling without dedicated lanes. From the perspective of eye tracking research and practice, this finding opens the possibility of developing automated or semi-automated tools for constructing narratives from gaze replay videos.  

\subsection{Theme 3: Ease of Use}
The above two themes involve applications that involve short term eye tracking data collection. 
For a training setting where durations were as small as 30-45 minutes, an experimenter took care of setting up and calibrating the equipment, and there were no auxiliary tasks to be performed, participants were not discomfited by the addition of eye tracking technology. \textit{Participant 6: All I did was put the glasses on my head. That’s all I did, he [gestures to researcher] did everything else.} \textit{Participant 9: Well, basically, all I did was wore it so, you know I didn't know how to, you know, know how to tap the phone and have you looked at the white and the one, three of it was I was like this. [describing calibration process]. But I think it’s designed pretty good whoever's working on it. And I gave it a thumbs up. I really do. I really do.}

Participants' responses to the NASA TLX reflected a corresponding low level of self-reported mental demand ($\mu$=1.9), physical demand ($\mu$=0.9), temporal demand ($\mu$=0.8), effort ($\mu$=0.6) and frustration ($\mu$=1.3) with possible values ranging from a minimum of $0$ to a maximum value of $10$. 
However, as they started thinking through the many ways in which this technology could become part of their routine driving work, they brought up possible burdens that might get imposed on them. This theme deals with the possibility of a well-intentioned technology becoming burdensome to the end user because it increases the quantity of their work or decreases the quality of their work. For example, transit drivers already do a number of things in addition to the task of driving: greeting customers, making announcements, securing wheelchairs, keeping in contact with the head office via radio. Needing to calibrate an eye tracker every time they got into their seat would add yet another task to what they have to do.

Drivers adjust the settings of the driver's seat and cab settings to be comfortable to them. Any new technology added to their cab creates one more parameter that they need to spend effort adjusting on a daily basis. Discomfort extends to what they wear on their person. Drivers in Florida typically wear sunglasses all year round. If they have to drive without their shades because they are required by management to do so for the purposes of eye tracking, this would increase their on-job discomfort. \textit{Participant 8: I'm not saying they're not cool. I'm just saying that they're not comfortable, and one of the key factors for a driver is to be comfortable at all times. I don’t know if ya'll noticed how much effort I put into getting my seat and my mirrors and everything in exactly the right spot. Because if you're thinking about the fact that you're not comfortable, you're not thinking about something you should be thinking about.}
Distraction must also be considered. For example, drivers have complained in the past that advanced driver assistance technology (ADAS) becomes distracting when there are too many alerts. Eye tracking technology could become distracting because of its form factor involving glasses, a cable, and a recording device. \textit{Participant 10: The only thing that kind of aggravate me at first was the cord, because like, I'm not used to something like you know, in my space.}

The drivers noted specific sources of discomfort that could be addressed relatively easily, for example, via an adjustable nose piece and wireless recording. \textit{Participant 9: The glasses could be a little looser because I got a fat face.} 


\textbf{Implications} The implications of these findings for eye tracking research and practice are that it is critical to design the form factor, weight, etc. in a way that the user is not discomfited by long duration use. Participatory design methods could be leveraged to create special purpose eye trackers for specific work contexts. Another implication is that to develop eye trackers that would work for users wearing eye glasses and consider novel designs such as clip-ons for commodity sunglasses.
\subsection{Theme 4: Accident Analysis} 
This theme highlights the perception of transit drivers that eye tracking technology could provide ``objective'' data for assessing who is at fault in case of an accident. Transit drivers felt that when an incident happens, there is often no way to resolve it beyond a ``he-said-she-said'' argument. Just as camera feed from roadway intersections help resolve whether a driver really jumped a signal and should be penalized, the gaze replay video from an eye tracker can be of benefit to the driver when they are not at fault. Participants also recognized that the opposite also holds: that the same data is beneficial to the management if the driver was indeed at fault. 
\textit{Participant 10: Like if, if there was a way to design it into, say into a windshield to see where the eyes are going, you can use that in investigating an accident to either what was clear or not clear. Like for a driver as, “Hey, wait, you weren't looking where you need to be looking.” Or, you know what, they were doing all the right things, this was not their fault.}

\textbf{Implications} As eye tracking researchers, it would be important to note that the end user perception is that eye tracking data is ``objective'' and ``absolute truth'' in case of conflicts. However, as we know, this data is subject to measurement error, calibration related error, drift, etc. It might therefore be useful to consider what types of training needs to be designed for transportation agencies to communicate the limits of eye tracking technology while also understanding its benefits. Another possible future work direction would be to create visualizations for these different types of errors so that lay users can appropriately modulate their conclusions. 

\subsection{Theme 5: Fear of Replacement/Loss of Livelihood}
While Theme 1 (Training) and Theme 2 (Tool for Communication) dealt with use cases where a small data was collected for a short period of time at pre-specified points, Theme 3 (Accident Analysis) brings up the idea of continuous monitoring on an ongoing basis. Participants brought up concerns about what this data could be used for beyond accident analysis both of which is an example application that improves the quality of work for the driver. However, once large amounts of data are collected into databases, it could be used in ways that are not perceived by drivers as being in their favor. For example, it could be used to train an AI who would then take away their jobs. \textit{Participant 8: I think the technology is good as long as you're not going to use it to design an AI that replaces my job.} 

\textbf{Implications} Consistent with discussions in other application areas for eye tracking data and more generally, the use of large datasets to train AI, our findings indicate that it would be important to carefully consider the persistence of data collected for an agreed upon purpose and 
 whether it can be used for another purpose down the line. It would be important that every party that has access to the recorded gaze data and accompanying video agrees to what the data will be used for and what it will not be used for.

\section{Conclusion and Discussion}
We have presented findings from a novel naturalistic field study which sought to elicit transit bus operators' perceptions of eye tracking technology. Though there is active work in our research community on leveraging eye tracking technology for various transportation contexts including training~\cite{krejtz:etra:2023} and safety~\cite{kubler:etra:2021}, there has not been any investigation thus far into how end users perceive this technology. Considering user perceptions early on allows for smoother technology adoption. It can surface both concerns and novel applications from the end user perspective. However, eliciting user perceptions requires qualitative research methodology involving interviews and focus groups. For eye tracking technology in particular, this methodology still needs to be developed because, unlike wearable smart watches or smart phones, it is not something that is currently used in everyday life. Our work is a first step in this direction. By providing a mixed methods protocol as well as illustrating novel concerns and applications surfaced through this protocol, we hope to stimulate research in this direction, especially in situations involving imbalanced power dynamics, for example, workers facing employer mandated roll out of eye tracking technology, students being eye tracked for personalized learning, and drivers being required to turn on eye tracking for reduced auto insurance rates. 
\bibliographystyle{ACM-Reference-Format}
\bibliography{Bibliography/fixed}
\end{document}